\documentclass
      [aip,jcp,preprint,amsmath,amssymb,superscriptaddress,groupedaddress]{revtex4-1}

\usepackage{graphicx} 

\input epsf
\input rotate

\DeclareMathOperator\erf{erf}

\def\bea{\begin{eqnarray}}
\def\eea{\end{eqnarray}}
\def\ben{\begin{equation}}
\def\een{\end{equation}}
\def\benu{\begin{enumerate}}
\def\enu{\end{enumerate}}

\def\lsim {\ifmmode {\buildrel<\over\sim}}

\def\sss{\scriptscriptstyle\rm}

\def\1var{(\bx_1...\bx\N)}

\def\br{{\bf r}}

\def\b1{{\bf 1}}
\def\bx{{x}}

\def\xc{_{\sss XC}}

\def\N{_{\sss N}}
\def\H{_{\sss H}}

\def\ext{_{\rm ext}}

\def\sph_int{ {\int d^3 r}}

\def\infintd3r{ \int_{-\infty}^\infty d^3r\,}
\def\intd3r{ \int d^3r\,}

\def\laplace1d{\frac{d^2}{dx^2}}
\def\plaplace1d{\frac{d^2}{d{x'}^2}}

\def\padr2{\frac{\partial^2}{\partial r^2}}

\def\N{{\cal N}}
\def\a{{\alpha}}
\def\b{{\beta}}

\begin{document}

\title{Fragment-based Treatment of Delocalization and Static Correlation
Errors in Density-Functional Theory}
\author{Jonathan Nafziger} 
\affiliation{Department of Chemistry, Purdue University, 560 Oval Dr., West Lafayette IN 47907, USA}

\author{Adam Wasserman}
\email[Corresponding Author: ]{awasser@purdue.edu}
\affiliation{Department of Chemistry, Purdue University, 560 Oval Dr., West Lafayette IN 47907, USA}
\affiliation{Department of Physics and Astronomy, Purdue University, 525 Northwestern Ave., West Lafayette, IN 47907, USA}


\begin{abstract}

One of the most important open challenges in modern Kohn-Sham (KS) density-functional theory (DFT) is the correct treatment of systems involving fractional electron charges and spins. Approximate exchange-correlation (XC) functionals struggle with such systems, leading to pervasive delocalization and static correlation errors. We demonstrate how these errors, which plague density-functional calculations of bond-stretching processes, can be avoided by employing the alternative framework of partition density-functional theory (PDFT) even with simple local and semi-local functionals for the fragments. Our method is illustrated with explicit calculations on simple systems exhibiting delocalization and static-correlation errors, stretched H$_2^+$, H$_2$, He$_2^+$, Li$_2^+$, and Li$_2$. In all these cases, our method leads to greatly improved dissociation-energy curves.  The effective KS potential corresponding to our self-consistent solutions display key features around the bond midpoint; these are known to be present in the exact KS potential,  but are absent from most approximate KS potentials and are essential for the correct description of electron dynamics.

\end{abstract} 

\maketitle

\section{Introduction}
Fifty years after its proposal, the Kohn-Sham (KS) prescription \cite{KS65} of density-functional theory (DFT) \cite{HK64} continues to be one of the most practical formulations of the many-electron problem in quantum chemistry and solid-state physics. Improving on the accuracy and efficiency of KS-DFT calculations is a constant and pressing goal for the electronic-structure community \cite{M04}, which works on understanding the sources of errors \cite{CMY08a,MCY08,MCY09,CMY08b}, developing new exchange-correlation (XC) functionals \cite{PBE96,TPSS03,ZT08}, and designing better and faster computational algorithms \cite{octopus}.

\begin{figure}[htp]
\includegraphics*[width=3.375in]{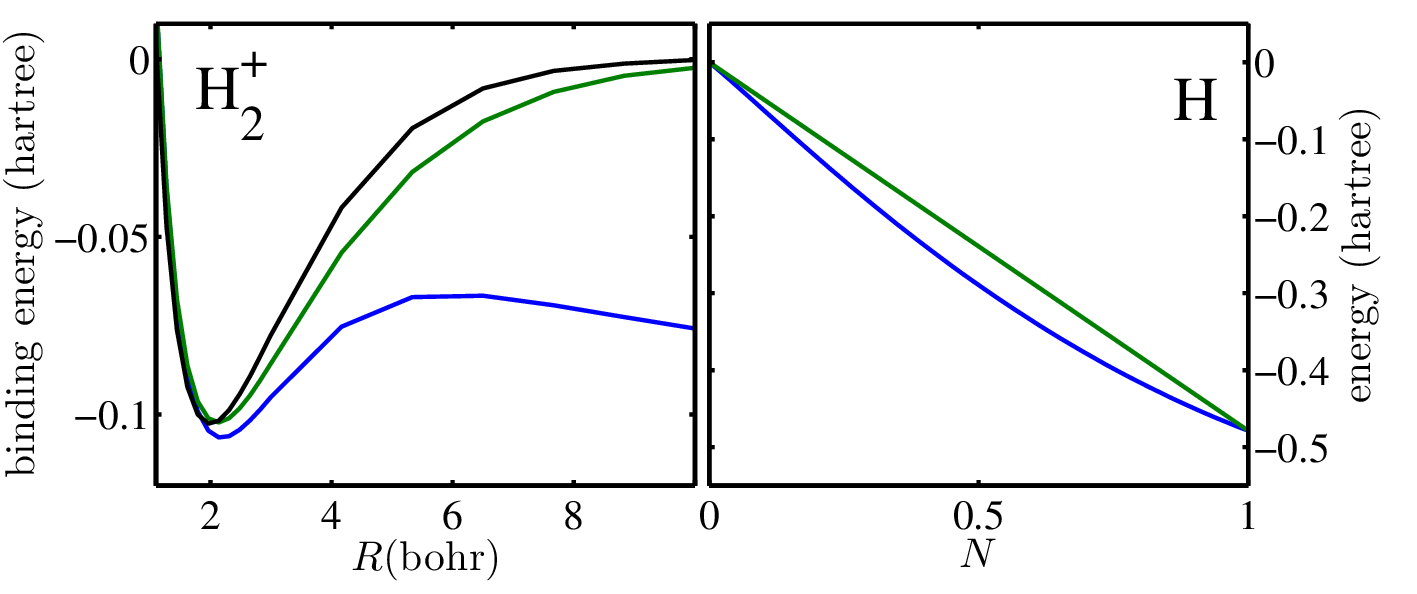}
\includegraphics*[width=3.375in]{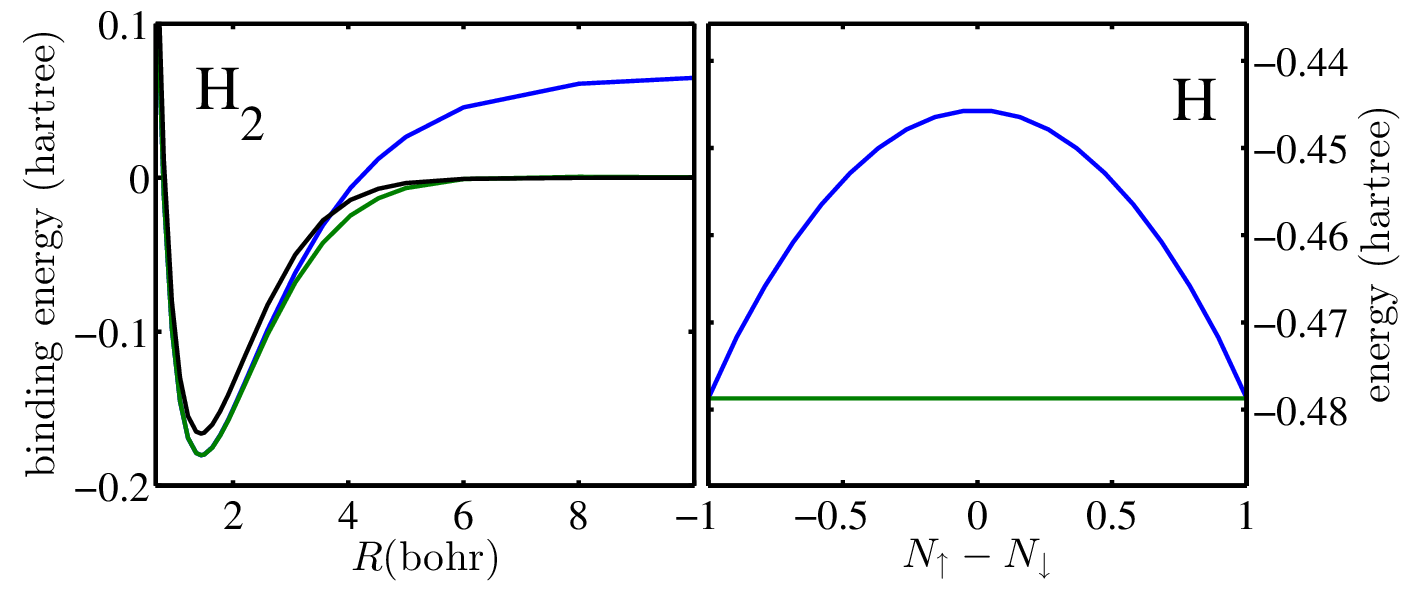}
\includegraphics*[width=3.375in]{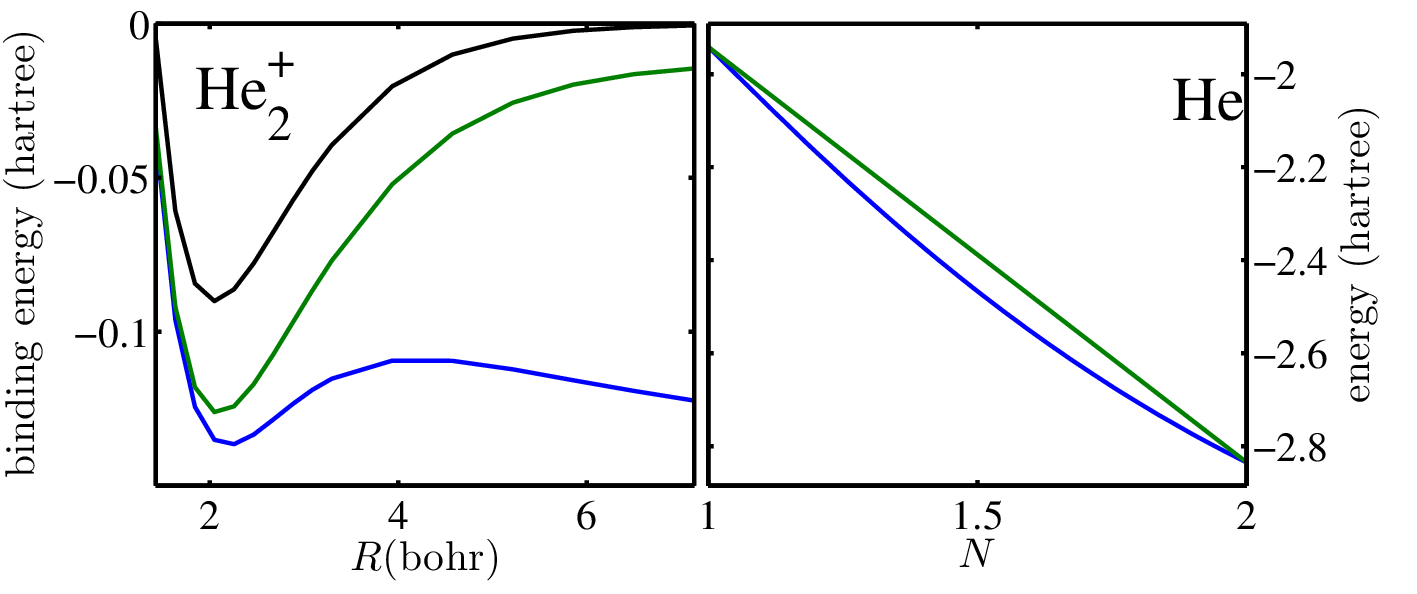}
\includegraphics*[width=3.375in]{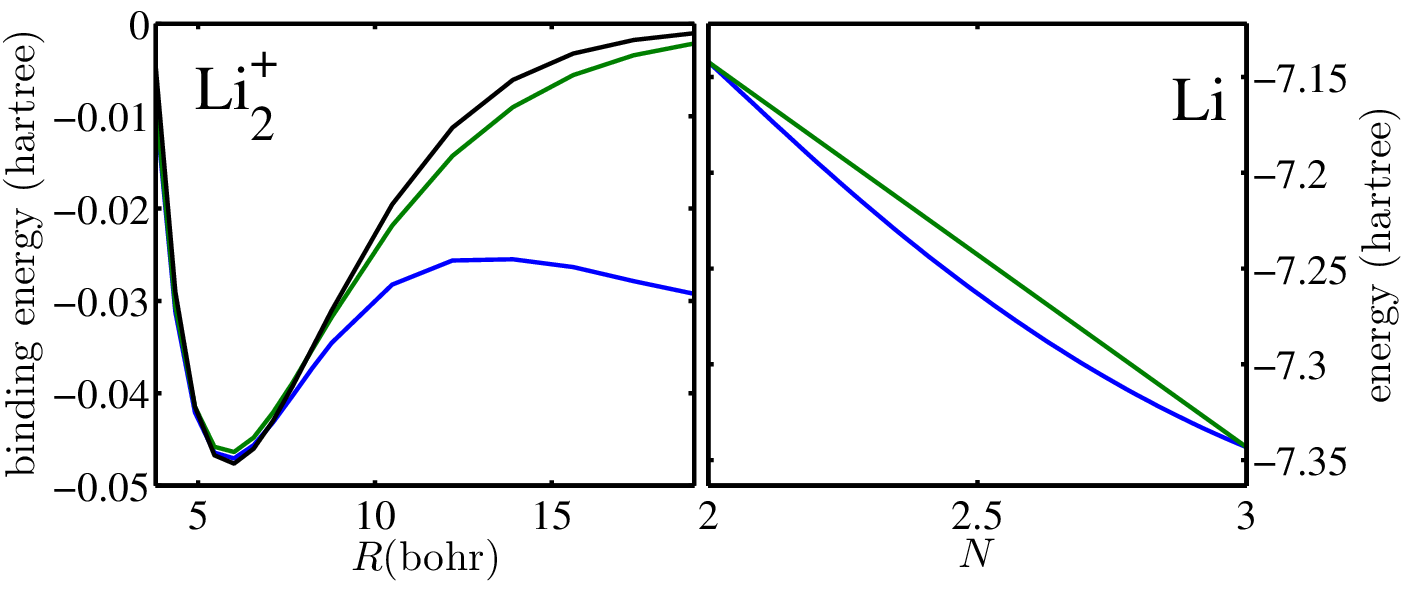}
\includegraphics*[width=3.375in]{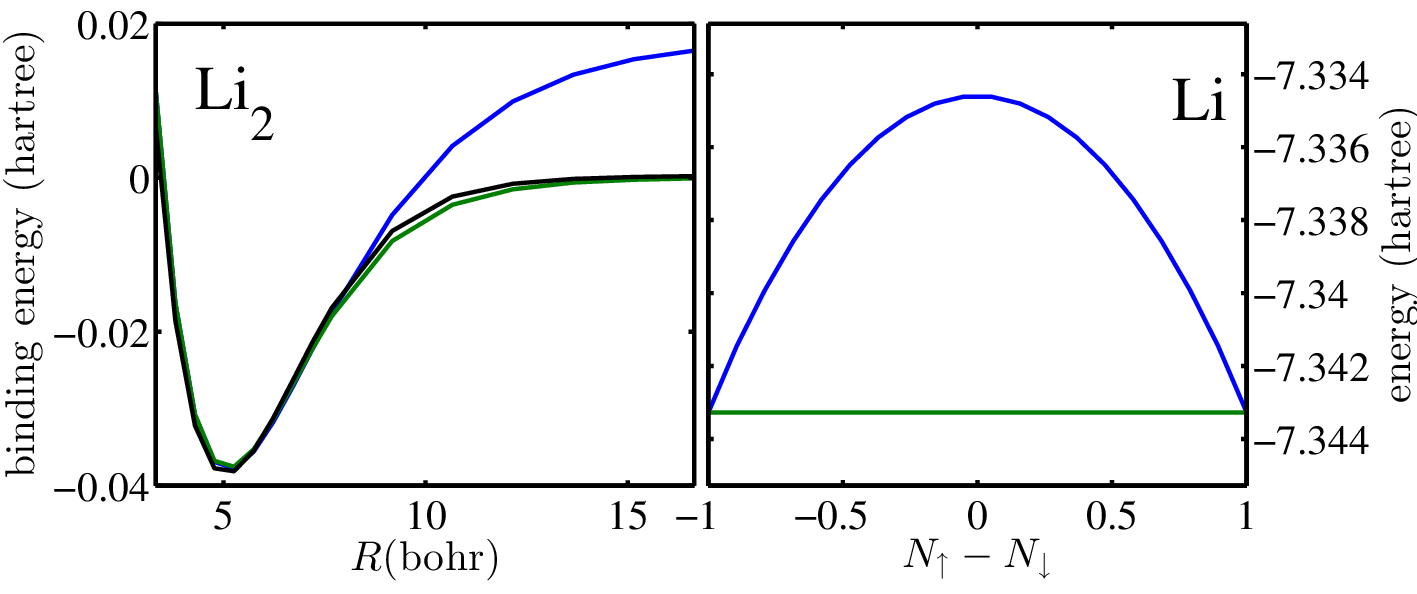}
\caption{Left column: Dissociation curves for five dimers.  In blue are the KS-LDA results, in green are the OA-LDA PDFT results and in black are CCSD/aug-cc-pVDZ calculations performed in NWChem\cite{nwchem} (except for H$_2^+$, which is simply a one-electron non-interacting calculations).  Right column: Energies of monomers for fractional charges and spins.  In blue are the KS-LDA energies evaluated for these fractional charges and spins and the green curves are the PDFT fragment energies.  The maximum difference between the blue and green curves in the right column account for half the difference between the blue and green curves in the dissociation limit (Left column).}
\label{fig:H2Li2He2p}
\end{figure}

Two open problems in DFT are the delocalization and static correlation errors of approximate functionals, arising from improper treatment of fractional charges and spins, respectively \cite{CMY08a,MCY08,MCY09,CMY08b}.  Delocalization errors cause underestimation of energies in dissociating molecular ions, chemical reaction barrier heights, charge-transfer excitations, band-gaps of semiconductors, as well as overestimation of binding energies of charge-transfer complexes and response to electric fields. Static correlation errors are responsible for the problems with degenerate and near-degenerate states, incorrect dissociation limit of neutral diatomics and poor treatment of strongly correlated systems.  The simplest systems that display these errors are stretched H$_2^+$, H$_2$, He$_2^+$, Li$_2^+$, and Li$_2$. Local and semi-local approximations to the exchange-correlation energy ($E\xc$) severely underestimate the dissociation energy of H$_2^+$, He$_2^+$ and Li$_2^+$ due to delocalization, and overestimate the dissociation energy of H$_2$ and Li$_2$ due to static correlation (See Figure \ref{fig:H2Li2He2p}).  

In this work we demonstrate that partition density-functional theory (PDFT) \cite{EBCW10} is a suitable framework to solve these problems. The {\em partition energy} of PDFT (denoted $E_p$, to be defined below) is amenable to simple approximations which can handle delocalized and statically-correlated electrons, greatly improving dissociation curves.  For example, Fig. \ref{fig:H2Li2He2p} displays the results we obtained by applying PDFT with the Local Density Approximation (LDA) and a simple ``Overlap Approximation" (OA) for $E_p$ (defined in Eq.\ref{e:OA}) as compared to standard KS-LDA results. We are not aware of approximate XC-functionals that yield similar accuracy for all these systems within standard KS-DFT.

PDFT allows a molecular calculation to be performed on individual fragments of a molecule rather than on the molecule as a whole. It is based on the density-partitioning scheme of refs. [\onlinecite{CW06}] and [\onlinecite{CW07}], and is nearly equivalent in practice to the formulation of embedding theory by Huang and Carter\cite{HC11} based on earlier work of Cortona \cite{C91} and Wesolowski and Warshel \cite{WW93} (see refs. [\onlinecite{JN2014}] and [\onlinecite{KSGP2015}] for recent reviews on subsystem-DFT). One critical difference, essential for this work, is our use of {\em ensemble} functionals to treat non-integer electron numbers and spins in fragments of molecules.  Kraisler and Kronik have also recently used ensemble-generalized functionals to solve issues with fractional charges in dissociation problems \cite{KK15}, but PDFT allows the use of these functionals at finite separations rather than being limited to completely isolated fragments.

\section{PDFT}
A detailed overview of PDFT may be found in ref. [\onlinecite{NW14}].  Here we only provide a brief summary.
The first step of PDFT is to partition the molecule into fragments.  This is done by dividing the nuclear potential into fragment potentials, $v_\a(\br)$.  This is the only choice the PDFT user makes in regards to the fragmentation and after this choice the fragment densities are uniquely determined\cite{CW06}.  We typically choose atom-based fragments, where each nucleus controls its own fragment, but any partition may be chosen as long as the sum of fragment potentials equals the total external nuclear potential. 

Each individual fragment calculation is a standard DFT calculation for each of the ensemble components of the ground-state density of $N_\a$ electrons in an effective potential.  We denote the $i^{\rm th}$ component of the $\a^{\rm th}$ fragment spin $\sigma$ density as $n_{i\a\sigma}(\br)$.  The number of electrons in each fragment's ensemble spin component, $N_{i\a\sigma}=\int{n_{i\a\sigma}(\br)d\br}$, will always be an integer number of electrons, but the total number of electrons of a given spin in a given fragment,
\begin{equation}
\label{e:Nalpha}
N_{\a\sigma} = \sum_if_{i\a}N_{i\a\sigma}
\end{equation}
will not neccesarily be an integer.  Here, the $f_{i\a}$ are the ensemble coefficients, which satisfy the sum rule, $\sum_if_{i\a}=1$.  The energy of these fragments is given by,
\begin{equation}
\label{e:Ealpha}
E_{\a} = \sum_if_{i\a}E_\a[n_{i\a\uparrow},n_{i\a\downarrow}]
\end{equation}
Here, the subscript $\alpha$ on the energy denotes that this is the energy corresponding to the $\{n_{i\a\sigma}\}$ in the external potential $v_\a(\br)$ rather than the total external potential.  The effective \emph{external} potential for each fragment is the sum of the fragment's potential, $v_\a(\br)$, and the partition potential, $v_{p\sigma}(\br)$.  The latter is a {\em global} quantity ensuring that the fragment calculations produce densities that sum to yield the correct molecular density while minimizing the sum of the fragment energies, $E_f$.  The partition potential enters formally as a lagrange multiplier constraining the fragment densities to equal the molecular density, but can be calculated as the functional derivative of $E_p$ with respect to the total density \cite{MW13}.

The partition energy, $E_p$, central to our work, is the difference between the total molecular energy, $E[n]$, and the sum of the fragment energies, $E_f=\sum_\a E_\a$.  As argued in ref. [\onlinecite{MW13}], the minimum value of $E_f$ with respect to variations of the $n_{i\a\sigma}$'s is a functional of the total density. Subtracting this quantity from the true ground-state energy yields $E_p[n]=E[n]-E_f[n]$, an implicit functional of the molecular density. We may also write $E_p$ as an explicit functional of the fragment densities: $E_p[\{n_{i\a\sigma}\}]=E[n_f]-E_f[\{n_{i\a\sigma}\}]$.  In the two-fragment case where each fragment has two ensemble components, $E_p$ can be divided into components and written out explicitly in terms of fragment densities:
\begin{equation}
\begin{aligned}
\label{e:Ep}
E_p[\{n_{i\a\sigma}\}] = T_s^{\rm nad}[\{n_{i\a\sigma}\}]+&V\ext^{\rm nad}[\{n_{i\a\sigma}\}]\\+ E\H^{\rm nad} [\{n_{i\a\sigma}\}] +&E\xc^{\rm nad}[\{n_{i\a\sigma}\}]~~,\\
\end{aligned}
\end{equation}
where each non-additive functional is,
\begin{equation}
\begin{aligned}
\label{e:Fnad}
F^{\rm nad}[\{n_{i\a\sigma}\}] \equiv F[n_{f\uparrow},n_{f\downarrow}] - \sum_{i,\a}{f_{i\a}F_\a[n_{i\a\uparrow},n_{i\a\downarrow}]}.
\end{aligned}
\end{equation}  
These are similar to the non-additive functionals of embedding theory \cite{HC11,C91,WW93} except that the functional values for each fragment are calculated from ensembles, rather than being evaluated on total fragment densities.  In practice, a choice of density-functional approximation (DFA) must be made for $E\xc$ and $E\xc^{\rm nad}$.  In addition, $T_s^{\rm nad}$ requires writing the non-interacting kinetic energy as a functional of the density.  Approximate kinetic energy functionals may be used \cite{WEW98}, although $T_s^{\rm nad}$ can also be obtained from an inversion of the sum of fragment densities as in ref. [\onlinecite{GAMM10}].  We use a similar inversion method for He$_2^+$, Li$_2^+$ and Li$_2$, and we use von Weizs\"{a}cker inversion for H$_2^+$ and H$_2$, since these systems have a single occupied orbital.  

For a given choice of XC functional, we may {\em exactly} reproduce the corresponding KS-DFT calculation as long as the same DFA is employed for both $E\xc^{\rm nad}$ and $E_f$ \cite{NWW11}.  We can also trivially reproduce a KS-DFT calculation by setting the number of fragments equal to one.  In these ways PDFT subsumes KS-DFT.  

\section{Approximating the Partition Energy Functional}
\begin{figure}[htp]
\includegraphics*[width=3.375in]{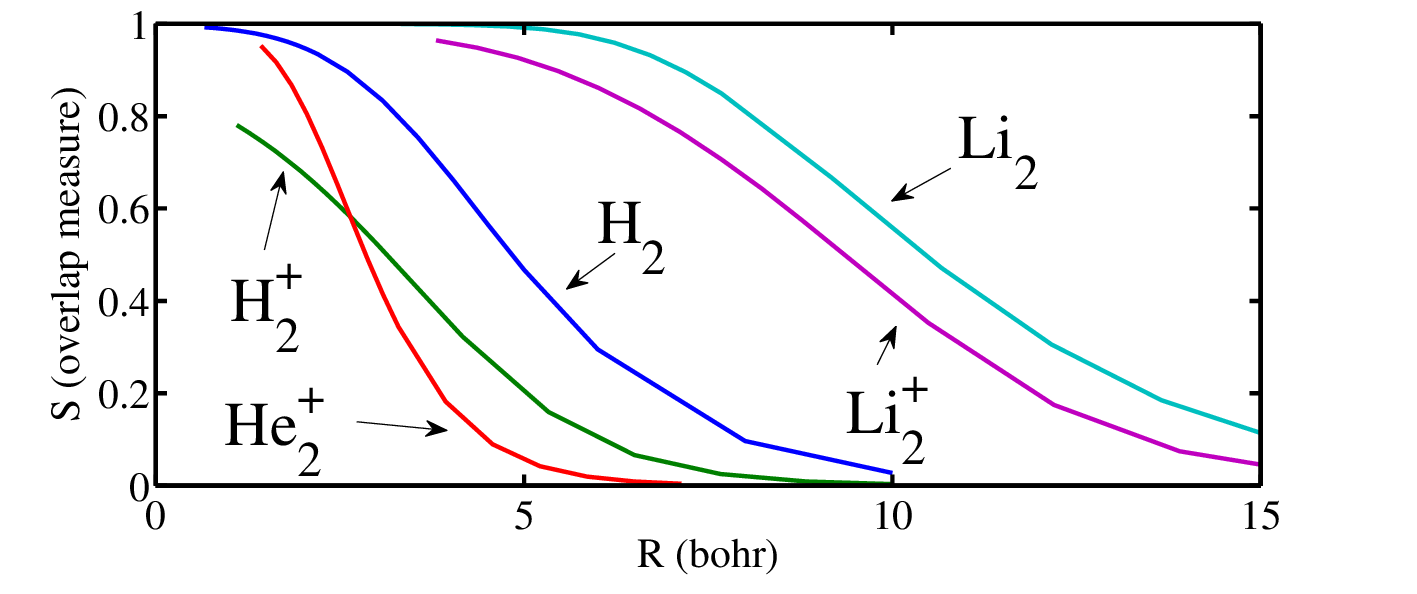}
\includegraphics*[width=3.375in]{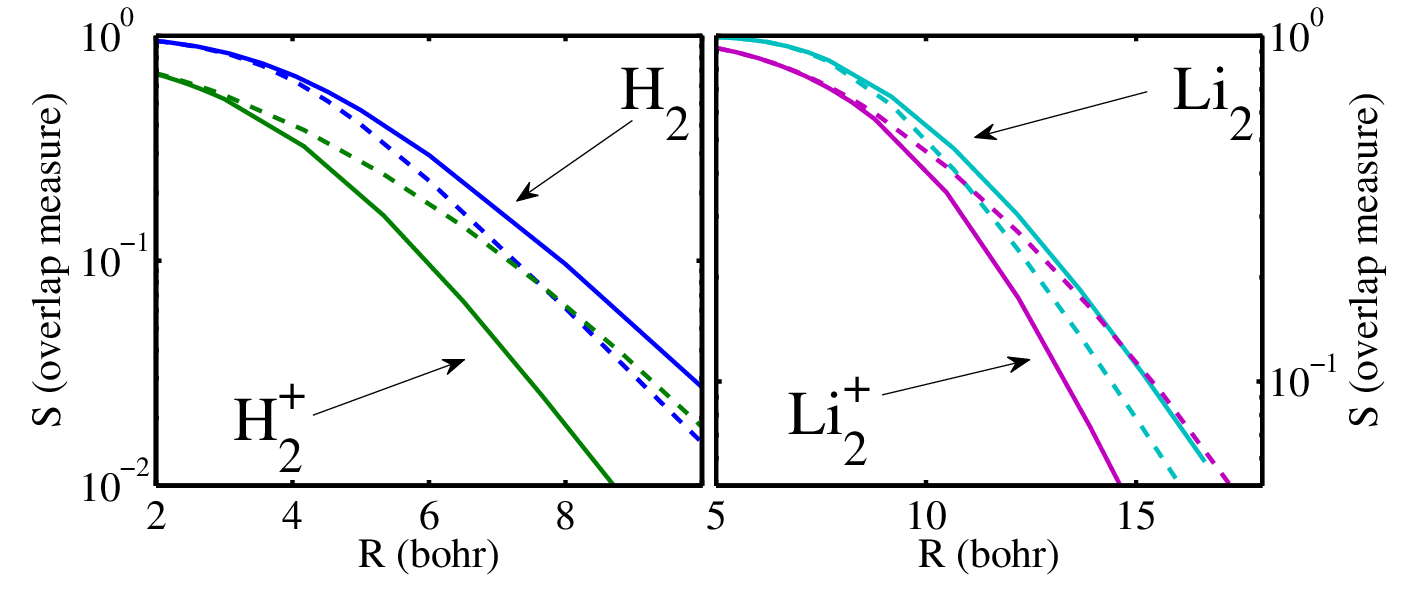}
\caption{Overlap measure of Eq. \ref{e:S} for various separtions.  Top: Overlap measure calculated on self consistent densities from standard PDFT calculations (using Eq. \ref{e:Ep}) which do not depend directly on the overlap measure.  The bottom two (logarithmic) plots compare the overlap measure evaluated for densities from PDFT calculations using Eq. \ref{e:Ep} (solid) and the OA, Eq. \ref{e:OA}, (dashed).  We see that when the OA is employed, the overlap measure is increased in the cases of H$_2^+$ and Li$_2^+$, while in thethe cases of H$_2$ and Li$_2$ the overlap measure is reduced.}
\label{fig:S}
\end{figure}

However, PDFT also goes beyond KS-DFT.  For example, the following ``Overlap Approximation" to the partition energy functional produces the results displayed in Fig. \ref{fig:H2Li2He2p} when used with LDA:
\ben
E_p^{\rm OA}=T_s^{\rm nad} + V\ext^{\rm nad} + E_{\rm H}^{\rm nad} + S E_{\rm xc}^{\rm nad} + (1-S)\Delta E_{\rm H}^{\rm nad}~,
\label{e:OA}
\een
where $\Delta E_{\rm H}^{\rm nad}$ is a correction to the non-additive hartree (to be discussed after its definition in Eq. \ref{e:Ehcorr}) , valid at larger separations, and $S$ is a functional of the fragment densities defined by:
\ben
S[n_A,n_B]=\erf(C \int\sqrt{n_A(\br)n_B(\br)} d\br)~.
\label{e:S}
\een
The densities, $n_A(\br)$ and $n_B(\br)$ refer to fragment densities summed over the ensemble and spin components.  The overlap measure, $S$, is designed to go to zero at infinite fragment separation and to one at equilibrium distances (reminiscent of the work of ref. [\onlinecite{MM04}]).  When $S=1$ the partition energy matches eq. \ref{e:Ep}, and thus the total energy will simply reproduce the standard KS energy for a given choice of XC functional.  As $S\rightarrow0$, the non-additive XC energy is turned off and a correction to the non-additive hartree, $\Delta E_{\rm H}^{\rm nad}$, is turned on.  There is one parameter $C$, which we have set to $C=2$ to yield the results in Fig. \ref{fig:H2Li2He2p}.  The values of the overlap measure, $S$, from these calculations are plotted in Fig. \ref{fig:S}.

Clearly, the separation of $E_p$ and $E_f$ opens opportunities for new approximations within a self-consistent framework.  In particular, when the error of a DFT calculation is due to fragmentation, as in bond-stretching, expressing $E_p$ as a functional of the set of fragment densities has the potential of fixing the error from its root. The physics of \emph{inter}-fragment interactions is contained in $E_p$ while that of \emph{intra}-fragment interactions is contained in $E_f$.
This is the main idea we wish to explore in the remainder of this paper. We first discuss a consequence of using different levels of approximation for $E_p$ and $E_f$.  As shown in ref. [\onlinecite{MW13}], the partition potential is determined from the chain rule (spin notation is supressed here and below for simplicity):
\ben
v_p({\bf r}) = \sum_\a \int d{\bf r'} v_{p,i\a}({\bf r'})Q_{i\a}({\bf r'},{\bf r})~,
\een
where the $\{i,\a\}^{\rm th}$ component of the partition potential is given by  
\ben
v_{p,i\a}({\bf r}) = \frac{\delta E_p }{ \delta n_{i\a}({\bf r})}
\een
and 
$
Q_{i\a}({\bf r'},{\bf r}) = \delta n_{i\a}({\bf r'}) / \delta n({\bf r})
$
satisfies the sum-rule:
$
\sum_{i,\a} Q_{i\a}({\bf r'},{\bf r}) = \delta({\bf r'}-{\bf r})
$.
As long as the same level of approximation is employed for $E_p$ and $E_f$, then at convergence $v_{p,i\a}({\bf r})=v_{p,j\b}({\bf r}) \;\forall \a,\b, i, j$, so the choice of $Q_{i\a}$ is inconsequential provided the sum-rule is satisfied.  When different levels of approximation are used for $E_p$ and $E_f$, however, the $v_{p,\a}({\bf r})$ are not necesarily identical at convergence, and it becomes critical to specify the approximation being used for the $Q_\a$.  Future work will need to establish the effect of different approximations for $Q$ on final energies and densities.  Throughout the present work, we employ the Local-Q approximation suggested in ref. [\onlinecite{MW13}]:
\begin{equation}
Q_{i\alpha}(\br,\br') =  \delta(\br-\br')\frac{n_{i\a}(\br)}{n_f(\br')}
\end{equation}

We now have all of the neccesary tools to perform PDFT calculations with separate approximations for $E_f$ and $E_p$.  We implemented PDFT on a real-space prolate spheroidal grid, following the work of Becke and other workers \cite{B82,MKK09,KLS96,LPS83,GKG97}, and found XC potentials and energies through use of the Libxc library \cite{libxc}.  We validated the code through calculations on H$_2^+$, H$_2$, and Li$_2$ at equilibrium geometries where our code yields the same energies to within $10^{-7}$ hartrees for for both PDFT (Using Eq. \ref{e:Ep}) and standard KS-DFT calculations.
(see table \ref{tab:validation} for a sample of such comparisons).  NWChem was used for reference CCSD/aug-cc-pVDZ calculations\cite{nwchem}.  We now look at the delocalization and static-correlation errors from the point of view of PDFT, and demonstrate our proposed solutions.
\begin{table*}[htp]
\begin{tabular*}{\textwidth}{@{\extracolsep{\fill} } c c c c c c }
\hline\hline
 \multicolumn{2}{c}{Li$_2$ LDA @ R = 5.120 bohr} & \multicolumn{2}{c}{H$_2$ LDA @ R = 1.446 bohr} & \multicolumn{2}{c}{H$_2^+$ Exact @ R = 2.0 bohr}    \\ 
\cline{1-2} \cline{3-4} \cline{5-6}
 KS-DFT\cite{GKG97} & PDFT &  KS-DFT\cite{GKG97} &  PDFT &  KS-DFT\cite{MKK09} &  PDFT \\
\hline
 14.7245 & -14.724457 & -1.137692 & -1.1376923 & -0.6026342144(7) & -0.60263425
 \\
\hline\hline
\end{tabular*}
\caption{Comparison of total energies in hartree, for our PDFT code, and from benchmark KS-DFT calculations.}
\label{tab:validation}
\end{table*}

\section{Delocalization}
We first consider the accuracy of $E_p$ vs. $E_f$ in H$_2^+$, He$_2^+$ and Li$_2^+$. Since the Hamiltonian in these cases has inversion symmetry, and the total number of electrons in each case is odd, the correct ground-state density has fractional numbers of electrons on the left and right sides.  In the case of H$_2^+$ this means ``half an electron" on the left and ``half an electron" on the right, but the correct ground-state energy at infinite separation is that of an isolated hydrogen atom (-0.5 hartree). A correct size-consistent electronic-structure method must therefore assign an energy of -0.25 hartree to a hydrogen atom with {\em half} an electron. This same argument may be extended to dissociating hydrogen chains, resulting in the conclusion that the energy is a piecewise-linear function of electron number \cite{YZA00}.  In the other two cases (He$_2^+$ and Li$_2^+$) this indicates that the correct energy of a fragment at infinite separation is a linear interpolation between two electronic systems with integer number of electrons: one with one more electron than the other.  This is of course accomplished by the exact grand-canonical ensemble functional \cite{PPLB82}, but it is {\em not} accomplished by most approximate functionals, as can be seen in Fig.\ref{fig:H2Li2He2p}
for LDA \cite{D30,PW92}.  For H$_2^+$ the self-interaction error ${\rm SIE}=E\H[n]+E\xc[n]$ is a convex function of electron number $N$.  As a consequence, LDA underestimates the energy for half an electron in a hydrogen atom.  Two times this error is precisely $E\H(\infty)^{\rm nad}+ E\xc(\infty)^{\rm nad}$ in Eq.(\ref{e:Ep}), the LDA delocalization error of H$_2^+$ at infinite separation. The OA of Eq.\ref{e:OA} works by suppressing this error as $S(\infty)=0$ and reproduces the LDA at the equilibrium separation.

Because PDFT treats each fragment using an ensemble, the fragment calculation for the left or right half of stretched H$_2^+$ is  a linear interpolation between open shell calculations for zero and one electron.  For He$_2^+$ the two ensemble components contain $1$ and $2$ electrons, i.e. for fragment $A$: $N_{1A}=1$, $N_{2A}=2$, $f_{1A}=1/2$ and $f_{2A}=1/2$.  For Li$_2^+$ the ensemble components contain $2$ and $3$ electrons ($N_{1A}=2$, $N_{2A}=3$, $f_{1A}=1/2$ and $f_{2A}=1/2$).  The energies and densities are linear interpolations between these ensemble components.  We call this interpolation ensemble-LDA (ELDA), and plot the resulting curves in the right hand column of Fig. \ref{fig:H2Li2He2p}.  Even LDA provides reasonable approximations for $E_f$ because each fragment calculation is done for a well-localized density with an integer number of electrons.  The ensemble formulation then provides the correct scaling for the energy of each fragment with respect to number of electrons in that fragment.  Thus, overall, our conclusion is that $E_f$ is reasonably accurate and it is $E_p$ which is causing error in the dissociation limit.  We now explain how Eq. \ref{e:OA} corrects $E_p$.

While in the case of H$_2^+$ it is clear that $E\H(\infty)^{\rm nad}$ is entirely equal to the self-interaction error, the cases of He$_2^+$ and Li$_2^+$ must be treated with more care.  This is the reason for the $\Delta E_{\rm H}^{\rm nad}$ term in Eq. \ref{e:OA}.  This correction is defined as
\ben
\Delta E_{\rm H}^{\rm nad} \equiv \frac{1}{4} \sum_{\alpha\ne\beta}\sum_{i,j} g_{i,j} f_{i\alpha} \int\frac{n_{iA}(\br)n_{jB}(\br')}{|\br-\br'|}d\br d\br' - E_{\rm H}^{\rm nad}~, 
\label{e:Ehcorr}
\een
where $g_{i,j} = 0$ if $N_{i\a}+N_{j\b}\ne N$ and $g_{i,j} = 1$ if $N_{i\a}+N_{j\b}=N$.  This makes it so that the ensemble component with one less electron on one fragment will only interact with the ensemble component with one more electron on the other fragment and vice-versa.  In the case of H$_2^+$ this means that the correction term is simply $- E_{\rm H}^{\rm nad}$ because the lower ensemble component has no electrons.  In the case of Li$_2^+$, for each fragment, one ensemble component has two electrons and the other has three electrons.  The first term of this interaction will simply be a fifty percent mixture of the electrostatic interaction between the two-electron component density on one fragment with the three-electron component density on the other side and vice-versa.  For cases where the ensemble components have the same total number of electrons such as H$_2$ and Li$_2$, the first term is exactly equal to $E_{\rm H}^{\rm nad}$ and this correction has no effect. 

This aproximation was inspired in part by range-separated hybrid (RSH) functionals \cite{BLS10}.  In RSH functionals, a larger portion of exact exchange is included in long-range interactions to improve accuracy.  The distinction between long-range and short-range is made by a tunable parameter.  In our case we also attempt to use an improved approximation for the long-range interaction, but our distinction between long and short range is contained in the separation of $E_f$ and $E_p$.  

\section{Static-Correlation}
We next see how this idea can be applied successfully to handle static correlation, taking H$_2$ as an example.  As in the H$_2^+$ case, we first consider the dissociation products of H$_2$: two isolated hydrogen atoms, with a total energy of -1.0 hartree.  However, the molecular calculation is spin-neutral, and it remains spin-neutral throughout dissociation due to inversion symmetry.  Therefore, each dissociating hydrogen atom has an electron which is ``half spin up"  and ``half spin down".  The exact functional assigns an energy to this fragment equal to that of a spin-up electron in a hydrogen atom.  This is known as the constancy condition \cite{CMY08b}.  However, approximate functionals do not show this behaviour and typically overestimate the energy of a system with fractional spins.  This overestimation exactly matches the static correlation error of dissociated H$_2$, and is given by $E\xc^{\rm nad}(\infty)$.  Once again, Eq.\ref{e:OA} works by suppressing this error as $S(\infty)=0$.

Each fragment in an H$_2$ PDFT calculation contains one electron, but the energies and spin-densities are considered to be ensembles of a spin-up and a spin-down electron, i.e. $N_{1A\uparrow}=1$, $N_{1A\downarrow}=0$, $N_{2A\uparrow}=0$, $N_{2A\downarrow}=1$, $f_{1A}=1/2$ and $f_{2A}=1/2$.  The energies and densities are then linear interpolations between a spin-up ensemble component and a spin-down ensemble component.  The case of Li$_2$ is similar.  The dissociation products are two isolated Li atoms.  The ensembles in a Li fragment within PDFT consist of two components: one with two spin-up electrons and one spin-down electron and the other with one spin-up electron and two spin-down electrons ($N_{1A\uparrow}=1$, $N_{1A\downarrow}=2$, $N_{2A\uparrow}=2$, $N_{2A\downarrow}=1$).  These two cases are degenerate so the fragment energies satisfy the constancy condition.  The energies and densities of these fragments are linear interpolations between these ensemble components.  As in the H$_2$ case, $E_f$ is accurate with standard DFA's and we only need to improve $E_p$.  

The OA of Eq. \ref{e:OA} works by imposing size-consistency on the partition energy:  at infinite separation $E_p$ must vanish.  For H$_2$ and Li$_2$ the only part of $E_p$ which does not go to zero is the $E\xc^{\rm nad}$ term.  Thus, the OA suprresses it through multiplication by $S$.

\section{Peak in the KS potential}

It is well known that the KS potential for stretched H$_2$ develops a peak at the bond midplane \cite{BBS89,LB94a,GLB95,GB96,GB97,HTR09,TMM09}. This exact feature of $v_s(\br)$, is essential for the correct description of dissociation and electron dynamics within KS-DFT\cite{EFRM12,FERM13}.  While certain sophisticated XC functionals such as those based on the random phase approximation can reproduce the peak\cite{HRG2012}, it is absent from most approximate DFA's.  It is clear from Fig. \ref{fig:H2Li2He2p} that the OA has greatly improved the dissociation energy for H$_2$, but we may also explore whether the OA can reproduce this peak in the XC potential.  We can derive the molecular XC potential corresponding to a PDFT calculation through the functional derivative of the XC energy, which in the case of PDFT can be broken into fragment pieces and non-additive pieces.
\begin{equation}
\begin{aligned}
v_{{\rm xc},\sigma}({\bf r}) &= \frac{\delta E\xc[n_\uparrow,n_\downarrow]}{\delta n_\sigma({\bf r})} 
\\&=\frac{\delta E_f^{\rm XC}[n_\uparrow,n_\downarrow]}{\delta n_\sigma({\bf r})}
+\frac{\delta E_p^{\rm XC}[n_\uparrow,n_\downarrow]}{\delta n_\sigma({\bf r})}\
\end{aligned}
\label{eq:vxc_component}
\end{equation}
\begin{figure}[htp]
\includegraphics*[width=3.375in]{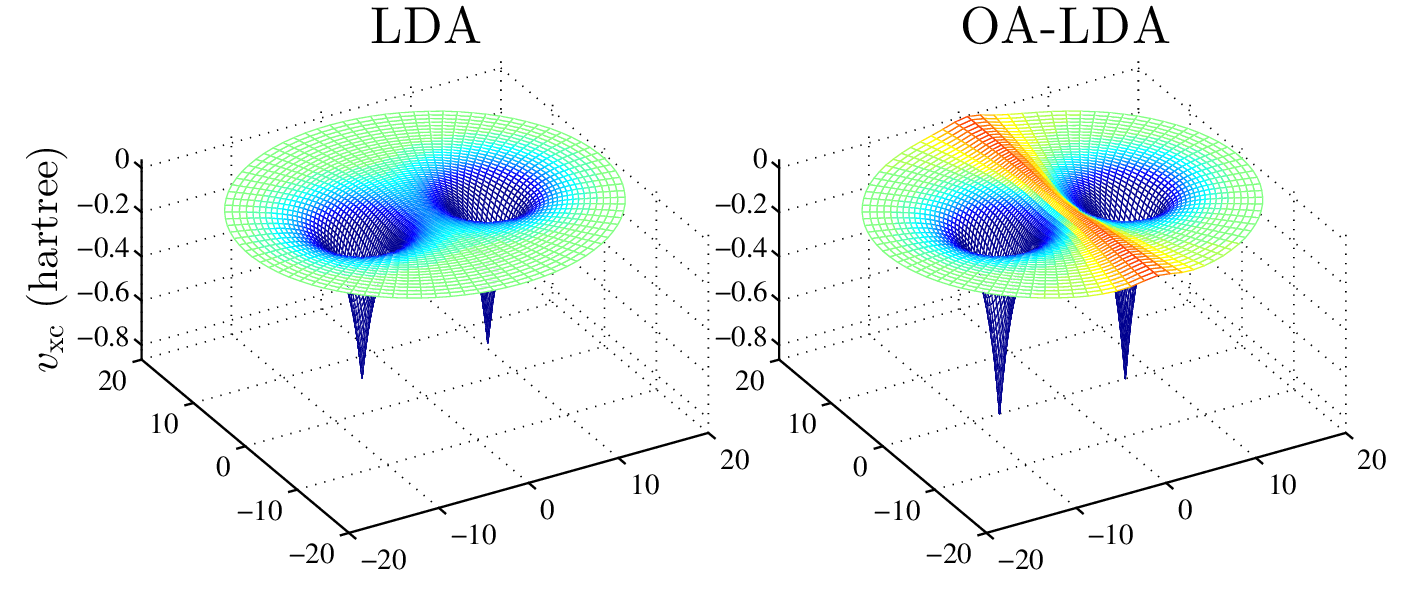}
\caption{Effective XC-potentials for two PDFT H$_2$ calculations with R = 14 bohr.  The plots show an entire 2D plane along the bonding axis.  Nuclei are at $+7$ and $-7$ on the bond axis.  Vertical scale for potentials is in units of hartree and horizontal scale is in bohr. }
\label{fig:peak}
\end{figure}

\begin{figure}[htp]
\includegraphics*[width=3.375in]{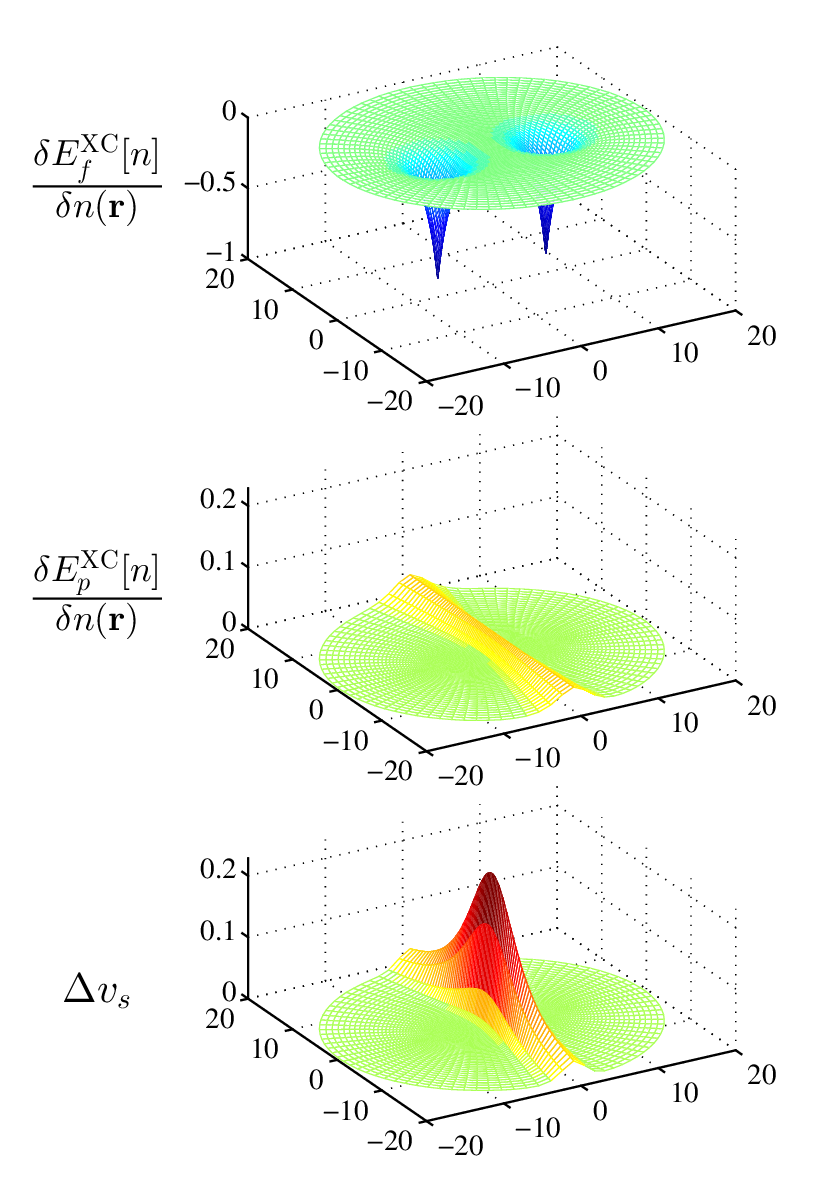}
\caption{Top and Middle: Components of the effective XC-potentials for OA-LDA H$_2$ calculations with R = 14 bohr, corresponding to the first and second terms of Eq. \ref{eq:vxc_component}.  Bottom: Peak in the KS potential from Eq. \ref{e:HTR}\cite{HTR09}.
Nuclei are at $+7$ and $-7$ bohr on the bond axis.  Vertical scale for potentials is in units of hartree and horizontal scale is in bohr. }
\label{fig:peak_component}
\end{figure}

Fig.\ref{fig:peak} compares this effective XC potential from two PDFT calculations on stretched H$_2$ (internuclear separation of 14 bohrs).  For the first we use the LDA in both $E_f$ and $E_p$.  For the second we use LDA in $E_f$ and OA-LDA for $E\xc^{\rm nad}$ in $E_p$.  We clearly see that the potential corresponding to the OA-LDA calculation has a peak in the bonding midplane.  Furthermore, we see that the peak comes entirely from the second term.  The two terms of Eq. \ref{eq:vxc_component} are plotted separately in Fig. \ref{fig:peak_component}.  

Helbig, Tokatly and Rubio (HTR) derived an exact form for $v_s(\br)$ for any two dissociated particles around the bonding midplane (Eq. 29 in their paper):
\ben
\begin{aligned}
\Delta v_s(\br) =& \frac{[\sqrt{n_A}\nabla\sqrt{n_B}-\sqrt{n_B}\nabla\sqrt{n_A}]^2}{2(n_A+n_B)^2}\\
&+\frac{I_An_A+I_Bn_B}{n_A+n_B}-I~,
\label{e:HTR}
\end{aligned}
\een
where, $n_A$ and $n_B$ are the dissociated atomic densities and $I_A$,$I_B$ and $I$ are the ionization potentials for the two fragments and the total system respectively.  In the case of homonuclear diatomics such as H$_2$, the second and third terms cancel leaving only the first term which produces a peak determined by the asymptotic behavior of $n_A$ and $n_B$.  HTR derived this expression using the fact that, at dissociation, the exact molecular density is exactly $n_m = n_A+n_B $, and neither the total nor fragment densities are represented by more than one orbital so the von Weizs\"{a}cker kinetic energy is exact.  In PDFT, the fragment densities $n_A$ and $n_B$ always sum to give the molecular density, so this expression may be evaluated at any finite separation.  In fact, Eq. \ref{e:HTR} can be identified as $-{\delta T_s^{\rm nad}}/{\delta n(\br)}$ from PDFT.

When we compare the peak produced by Eq. \ref{e:HTR} to the peak produced by the OA-LDA we see that its shape and size do not quite match.  The maximum value from the HTR expression is around $0.24$ hartree, while the maximum from the OA-LDA peak is $0.04$ hartree.  The HTR peak has its maximum at the bond axis and decreases farther away from the bond axis while the OA-LDA peak is flat.  Nevertheless, the OA-LDA peak is in the correct location and is localized to the same region in the bonding midplane.  Furthermore the fact that the HTR expression exactly matches $-{\delta T_s^{\rm nad}}/{\delta n(\br)}$ may help guide further development of functional forms for the OA.

\section{Concluding Remark}
The techniques described thus far are specific to homonuclear diatomics, but work is ongoing to extend these ideas to more general systems, including heteronuclear and multifragent systems. Our results suggest that local and semi-local density-functional approximations already do well for the localized fragments involved in the calculation of $E_f$ and attention needs to be placed on developing general approximations for $E_p$. This paper indicates that the path is worth taking, as even a simple approximation for $E_p$ can achieve via fragment calculations what sophisticated XC-functionals cannot via standard molecular calculations. 

{\em Acknowledgments:} 
We acknowledge valuable discussions with Mart\'{i}n Mosquera and Daniel Jensen. This work was supported by the Office of Basic Energy Sciences, U.S. Department of Energy, under grant No.DE-FG02-10ER16196.  A.W. also acknowledges support from the Alfred P. Sloan Foundation and the Camille Dreyfus Teacher-Scholar Awards Programs.

\bibliography{NW15}

\begin{thebibliography}{48}%
\makeatletter
\providecommand \@ifxundefined [1]{%
 \@ifx{#1\undefined}
}%
\providecommand \@ifnum [1]{%
 \ifnum #1\expandafter \@firstoftwo
 \else \expandafter \@secondoftwo
 \fi
}%
\providecommand \@ifx [1]{%
 \ifx #1\expandafter \@firstoftwo
 \else \expandafter \@secondoftwo
 \fi
}%
\providecommand \natexlab [1]{#1}%
\providecommand \enquote  [1]{``#1''}%
\providecommand \bibnamefont  [1]{#1}%
\providecommand \bibfnamefont [1]{#1}%
\providecommand \citenamefont [1]{#1}%
\providecommand \href@noop [0]{\@secondoftwo}%
\providecommand \href [0]{\begingroup \@sanitize@url \@href}%
\providecommand \@href[1]{\@@startlink{#1}\@@href}%
\providecommand \@@href[1]{\endgroup#1\@@endlink}%
\providecommand \@sanitize@url [0]{\catcode `\\12\catcode `\$12\catcode
  `\&12\catcode `\#12\catcode `\^12\catcode `\_12\catcode `\%12\relax}%
\providecommand \@@startlink[1]{}%
\providecommand \@@endlink[0]{}%
\providecommand \url  [0]{\begingroup\@sanitize@url \@url }%
\providecommand \@url [1]{\endgroup\@href {#1}{\urlprefix }}%
\providecommand \urlprefix  [0]{URL }%
\providecommand \Eprint [0]{\href }%
\providecommand \doibase [0]{http://dx.doi.org/}%
\providecommand \selectlanguage [0]{\@gobble}%
\providecommand \bibinfo  [0]{\@secondoftwo}%
\providecommand \bibfield  [0]{\@secondoftwo}%
\providecommand \translation [1]{[#1]}%
\providecommand \BibitemOpen [0]{}%
\providecommand \bibitemStop [0]{}%
\providecommand \bibitemNoStop [0]{.\EOS\space}%
\providecommand \EOS [0]{\spacefactor3000\relax}%
\providecommand \BibitemShut  [1]{\csname bibitem#1\endcsname}%
\let\auto@bib@innerbib\@empty
\bibitem [{\citenamefont {Kohn}\ and\ \citenamefont {Sham}(1965)}]{KS65}%
  \BibitemOpen
  \bibfield  {author} {\bibinfo {author} {\bibfnamefont {W.}~\bibnamefont
  {Kohn}}\ and\ \bibinfo {author} {\bibfnamefont {L.~J.}\ \bibnamefont
  {Sham}},\ }\href {\doibase 10.1103/PhysRev.140.A1133} {\bibfield  {journal}
  {\bibinfo  {journal} {Phys. Rev.}\ }\textbf {\bibinfo {volume} {140}},\
  \bibinfo {pages} {A1133} (\bibinfo {year} {1965})}\BibitemShut {NoStop}%
\bibitem [{\citenamefont {Hohenberg}\ and\ \citenamefont {Kohn}(1964)}]{HK64}%
  \BibitemOpen
  \bibfield  {author} {\bibinfo {author} {\bibfnamefont {P.}~\bibnamefont
  {Hohenberg}}\ and\ \bibinfo {author} {\bibfnamefont {W.}~\bibnamefont
  {Kohn}},\ }\href {\doibase 10.1103/PhysRev.136.B864} {\bibfield  {journal}
  {\bibinfo  {journal} {Phys. Rev.}\ }\textbf {\bibinfo {volume} {136}},\
  \bibinfo {pages} {B864} (\bibinfo {year} {1964})}\BibitemShut {NoStop}%
\bibitem [{\citenamefont {Martin}(2004)}]{M04}%
  \BibitemOpen
  \bibfield  {author} {\bibinfo {author} {\bibfnamefont {R.~M.}\ \bibnamefont
  {Martin}},\ }\href@noop {} {\emph {\bibinfo {title} {Electronic structure:
  basic theory and practical methods}}}\ (\bibinfo  {publisher} {Cambridge
  UP},\ \bibinfo {year} {2004})\BibitemShut {NoStop}%
\bibitem [{\citenamefont {Cohen}, \citenamefont {Mori-S\'{a}nchez},\ and\
  \citenamefont {Yang}(2008{\natexlab{a}})}]{CMY08a}%
  \BibitemOpen
  \bibfield  {author} {\bibinfo {author} {\bibfnamefont {A.~J.}\ \bibnamefont
  {Cohen}}, \bibinfo {author} {\bibfnamefont {P.}~\bibnamefont
  {Mori-S\'{a}nchez}}, \ and\ \bibinfo {author} {\bibfnamefont
  {W.}~\bibnamefont {Yang}},\ }\href@noop {} {\bibfield  {journal} {\bibinfo
  {journal} {Science}\ }\textbf {\bibinfo {volume} {321}},\ \bibinfo {pages}
  {792} (\bibinfo {year} {2008}{\natexlab{a}})}\BibitemShut {NoStop}%
\bibitem [{\citenamefont {Mori-S\'{a}nchez}, \citenamefont {Cohen},\ and\
  \citenamefont {Yang}(2008)}]{MCY08}%
  \BibitemOpen
  \bibfield  {author} {\bibinfo {author} {\bibfnamefont {P.}~\bibnamefont
  {Mori-S\'{a}nchez}}, \bibinfo {author} {\bibfnamefont {A.~J.}\ \bibnamefont
  {Cohen}}, \ and\ \bibinfo {author} {\bibfnamefont {W.}~\bibnamefont {Yang}},\
  }\href@noop {} {\bibfield  {journal} {\bibinfo  {journal} {Phys. Rev. Lett.}\
  }\textbf {\bibinfo {volume} {100}},\ \bibinfo {pages} {146401} (\bibinfo
  {year} {2008})}\BibitemShut {NoStop}%
\bibitem [{\citenamefont {Mori-S\'{a}nchez}, \citenamefont {Cohen},\ and\
  \citenamefont {Yang}(2009)}]{MCY09}%
  \BibitemOpen
  \bibfield  {author} {\bibinfo {author} {\bibfnamefont {P.}~\bibnamefont
  {Mori-S\'{a}nchez}}, \bibinfo {author} {\bibfnamefont {A.~J.}\ \bibnamefont
  {Cohen}}, \ and\ \bibinfo {author} {\bibfnamefont {W.}~\bibnamefont {Yang}},\
  }\href@noop {} {\bibfield  {journal} {\bibinfo  {journal} {Phys. Rev. Lett.}\
  }\textbf {\bibinfo {volume} {102}},\ \bibinfo {pages} {066403} (\bibinfo
  {year} {2009})}\BibitemShut {NoStop}%
\bibitem [{\citenamefont {Cohen}, \citenamefont {Mori-S\'{a}nchez},\ and\
  \citenamefont {Yang}(2008{\natexlab{b}})}]{CMY08b}%
  \BibitemOpen
  \bibfield  {author} {\bibinfo {author} {\bibfnamefont {A.~J.}\ \bibnamefont
  {Cohen}}, \bibinfo {author} {\bibfnamefont {P.}~\bibnamefont
  {Mori-S\'{a}nchez}}, \ and\ \bibinfo {author} {\bibfnamefont
  {W.}~\bibnamefont {Yang}},\ }\href@noop {} {\bibfield  {journal} {\bibinfo
  {journal} {J. Chem. Phys.}\ }\textbf {\bibinfo {volume} {129}},\ \bibinfo
  {pages} {121104} (\bibinfo {year} {2008}{\natexlab{b}})}\BibitemShut
  {NoStop}%
\bibitem [{\citenamefont {Perdew}, \citenamefont {Burke},\ and\ \citenamefont
  {Ernzerhof}(1996)}]{PBE96}%
  \BibitemOpen
  \bibfield  {author} {\bibinfo {author} {\bibfnamefont {J.~P.}\ \bibnamefont
  {Perdew}}, \bibinfo {author} {\bibfnamefont {K.}~\bibnamefont {Burke}}, \
  and\ \bibinfo {author} {\bibfnamefont {M.}~\bibnamefont {Ernzerhof}},\
  }\href@noop {} {\bibfield  {journal} {\bibinfo  {journal} {Phys. Rev. Lett.}\
  }\textbf {\bibinfo {volume} {77}},\ \bibinfo {pages} {3865} (\bibinfo {year}
  {1996})}\BibitemShut {NoStop}%
\bibitem [{\citenamefont {Tao}\ \emph {et~al.}(2003)\citenamefont {Tao},
  \citenamefont {Perdew}, \citenamefont {Staroverov},\ and\ \citenamefont
  {Scuseria}}]{TPSS03}%
  \BibitemOpen
  \bibfield  {author} {\bibinfo {author} {\bibfnamefont {J.}~\bibnamefont
  {Tao}}, \bibinfo {author} {\bibfnamefont {J.~P.}\ \bibnamefont {Perdew}},
  \bibinfo {author} {\bibfnamefont {V.~N.}\ \bibnamefont {Staroverov}}, \ and\
  \bibinfo {author} {\bibfnamefont {G.~E.}\ \bibnamefont {Scuseria}},\
  }\href@noop {} {\bibfield  {journal} {\bibinfo  {journal} {Phys. Rev. Lett.}\
  }\textbf {\bibinfo {volume} {91}},\ \bibinfo {pages} {146401} (\bibinfo
  {year} {2003})}\BibitemShut {NoStop}%
\bibitem [{\citenamefont {Zhao}\ and\ \citenamefont {Truhlar}(2008)}]{ZT08}%
  \BibitemOpen
  \bibfield  {author} {\bibinfo {author} {\bibfnamefont {Y.}~\bibnamefont
  {Zhao}}\ and\ \bibinfo {author} {\bibfnamefont {D.~G.}\ \bibnamefont
  {Truhlar}},\ }\href@noop {} {\bibfield  {journal} {\bibinfo  {journal}
  {Theor. Chem. Acc.}\ }\textbf {\bibinfo {volume} {120}},\ \bibinfo {pages}
  {215} (\bibinfo {year} {2008})}\BibitemShut {NoStop}%
\bibitem [{\citenamefont {Andrade}\ \emph {et~al.}(2012)\citenamefont
  {Andrade}, \citenamefont {Alberdi-Rodriguez}, \citenamefont {Strubbe},
  \citenamefont {Oliveira}, \citenamefont {Nogueira}, \citenamefont {Castro},
  \citenamefont {Muguerza}, \citenamefont {Arruabarrena}, \citenamefont
  {Louie}, \citenamefont {Aspuru-Guzik}, \citenamefont {Rubio},\ and\
  \citenamefont {Marques}}]{octopus}%
  \BibitemOpen
  \bibfield  {author} {\bibinfo {author} {\bibfnamefont {X.}~\bibnamefont
  {Andrade}}, \bibinfo {author} {\bibfnamefont {J.}~\bibnamefont
  {Alberdi-Rodriguez}}, \bibinfo {author} {\bibfnamefont {D.~A.}\ \bibnamefont
  {Strubbe}}, \bibinfo {author} {\bibfnamefont {M.~J.~T.}\ \bibnamefont
  {Oliveira}}, \bibinfo {author} {\bibfnamefont {F.}~\bibnamefont {Nogueira}},
  \bibinfo {author} {\bibfnamefont {A.}~\bibnamefont {Castro}}, \bibinfo
  {author} {\bibfnamefont {J.}~\bibnamefont {Muguerza}}, \bibinfo {author}
  {\bibfnamefont {A.}~\bibnamefont {Arruabarrena}}, \bibinfo {author}
  {\bibfnamefont {S.~G.}\ \bibnamefont {Louie}}, \bibinfo {author}
  {\bibfnamefont {A.}~\bibnamefont {Aspuru-Guzik}}, \bibinfo {author}
  {\bibfnamefont {A.}~\bibnamefont {Rubio}}, \ and\ \bibinfo {author}
  {\bibfnamefont {M.~A.~L.}\ \bibnamefont {Marques}},\ }\href {\doibase
  10.1088/0953-8984/24/23/233202} {\bibfield  {journal} {\bibinfo  {journal}
  {J. Phys.: Condens. Matter}\ }\textbf {\bibinfo {volume} {24}},\ \bibinfo
  {pages} {233202} (\bibinfo {year} {2012})}\BibitemShut {NoStop}%
\bibitem [{\citenamefont {Elliott}\ \emph {et~al.}(2010)\citenamefont
  {Elliott}, \citenamefont {Burke}, \citenamefont {Cohen},\ and\ \citenamefont
  {Wasserman}}]{EBCW10}%
  \BibitemOpen
  \bibfield  {author} {\bibinfo {author} {\bibfnamefont {P.}~\bibnamefont
  {Elliott}}, \bibinfo {author} {\bibfnamefont {K.}~\bibnamefont {Burke}},
  \bibinfo {author} {\bibfnamefont {M.~H.}\ \bibnamefont {Cohen}}, \ and\
  \bibinfo {author} {\bibfnamefont {A.}~\bibnamefont {Wasserman}},\ }\href@noop
  {} {\bibfield  {journal} {\bibinfo  {journal} {Phys. Rev. A}\ }\textbf
  {\bibinfo {volume} {82}},\ \bibinfo {pages} {024501} (\bibinfo {year}
  {2010})}\BibitemShut {NoStop}%
\bibitem [{\citenamefont {Cohen}\ and\ \citenamefont {Wasserman}(2006)}]{CW06}%
  \BibitemOpen
  \bibfield  {author} {\bibinfo {author} {\bibfnamefont {M.~H.}\ \bibnamefont
  {Cohen}}\ and\ \bibinfo {author} {\bibfnamefont {A.}~\bibnamefont
  {Wasserman}},\ }\href@noop {} {\bibfield  {journal} {\bibinfo  {journal} {J.
  Stat. Phys.}\ }\textbf {\bibinfo {volume} {125}},\ \bibinfo {pages} {1121}
  (\bibinfo {year} {2006})}\BibitemShut {NoStop}%
\bibitem [{\citenamefont {Cohen}\ and\ \citenamefont {Wasserman}(2007)}]{CW07}%
  \BibitemOpen
  \bibfield  {author} {\bibinfo {author} {\bibfnamefont {M.~H.}\ \bibnamefont
  {Cohen}}\ and\ \bibinfo {author} {\bibfnamefont {A.}~\bibnamefont
  {Wasserman}},\ }\href@noop {} {\bibfield  {journal} {\bibinfo  {journal} {J.
  Phys. Chem. A}\ }\textbf {\bibinfo {volume} {111}},\ \bibinfo {pages} {2229}
  (\bibinfo {year} {2007})}\BibitemShut {NoStop}%
\bibitem [{\citenamefont {Huang}\ and\ \citenamefont {Carter}(2011)}]{HC11}%
  \BibitemOpen
  \bibfield  {author} {\bibinfo {author} {\bibfnamefont {C.}~\bibnamefont
  {Huang}}\ and\ \bibinfo {author} {\bibfnamefont {E.~A.}\ \bibnamefont
  {Carter}},\ }\href@noop {} {\bibfield  {journal} {\bibinfo  {journal} {J.
  Chem. Phys.}\ }\textbf {\bibinfo {volume} {135}},\ \bibinfo {pages} {194104}
  (\bibinfo {year} {2011})}\BibitemShut {NoStop}%
\bibitem [{\citenamefont {Cortona}(1991)}]{C91}%
  \BibitemOpen
  \bibfield  {author} {\bibinfo {author} {\bibfnamefont {P.}~\bibnamefont
  {Cortona}},\ }\href@noop {} {\bibfield  {journal} {\bibinfo  {journal} {Phys.
  Rev. B}\ }\textbf {\bibinfo {volume} {44}},\ \bibinfo {pages} {8454}
  (\bibinfo {year} {1991})}\BibitemShut {NoStop}%
\bibitem [{\citenamefont {Wesolowski}\ and\ \citenamefont
  {Warshel}(1993)}]{WW93}%
  \BibitemOpen
  \bibfield  {author} {\bibinfo {author} {\bibfnamefont {T.~A.}\ \bibnamefont
  {Wesolowski}}\ and\ \bibinfo {author} {\bibfnamefont {A.}~\bibnamefont
  {Warshel}},\ }\href@noop {} {\bibfield  {journal} {\bibinfo  {journal} {J.
  Phys. Chem.}\ }\textbf {\bibinfo {volume} {97}},\ \bibinfo {pages} {8050}
  (\bibinfo {year} {1993})}\BibitemShut {NoStop}%
\bibitem [{\citenamefont {Jacob}\ and\ \citenamefont
  {Neugebauer}(2014)}]{JN2014}%
  \BibitemOpen
  \bibfield  {author} {\bibinfo {author} {\bibfnamefont {C.~R.}\ \bibnamefont
  {Jacob}}\ and\ \bibinfo {author} {\bibfnamefont {J.}~\bibnamefont
  {Neugebauer}},\ }\href@noop {} {\bibfield  {journal} {\bibinfo  {journal}
  {Wiley Interdisciplinary Reviews: Computational Molecular Science}\ }\textbf
  {\bibinfo {volume} {4}},\ \bibinfo {pages} {325} (\bibinfo {year}
  {2014})}\BibitemShut {NoStop}%
\bibitem [{\citenamefont {Krishtal}\ \emph {et~al.}(2015)\citenamefont
  {Krishtal}, \citenamefont {Sinha}, \citenamefont {Genova},\ and\
  \citenamefont {Pavanello}}]{KSGP2015}%
  \BibitemOpen
  \bibfield  {author} {\bibinfo {author} {\bibfnamefont {A.}~\bibnamefont
  {Krishtal}}, \bibinfo {author} {\bibfnamefont {D.}~\bibnamefont {Sinha}},
  \bibinfo {author} {\bibfnamefont {A.}~\bibnamefont {Genova}}, \ and\ \bibinfo
  {author} {\bibfnamefont {M.}~\bibnamefont {Pavanello}},\ }\href@noop {}
  {\bibfield  {journal} {\bibinfo  {journal} {Journal of Physics: Condensed
  Matter}\ }\textbf {\bibinfo {volume} {27}},\ \bibinfo {pages} {183202}
  (\bibinfo {year} {2015})}\BibitemShut {NoStop}%
\bibitem [{\citenamefont {Kraisler}\ and\ \citenamefont {Kronik}(2015)}]{KK15}%
  \BibitemOpen
  \bibfield  {author} {\bibinfo {author} {\bibfnamefont {E.}~\bibnamefont
  {Kraisler}}\ and\ \bibinfo {author} {\bibfnamefont {L.}~\bibnamefont
  {Kronik}},\ }\href@noop {} {\bibfield  {journal} {\bibinfo  {journal}
  {Physical Review A}\ }\textbf {\bibinfo {volume} {91}},\ \bibinfo {pages}
  {032504} (\bibinfo {year} {2015})}\BibitemShut {NoStop}%
\bibitem [{\citenamefont {Valiev}\ \emph {et~al.}(2010)\citenamefont {Valiev},
  \citenamefont {Bylaska}, \citenamefont {Govind}, \citenamefont {Kowalski},
  \citenamefont {Straatsma}, \citenamefont {Van~Dam}, \citenamefont {Wang},
  \citenamefont {Nieplocha}, \citenamefont {Apra}, \citenamefont {Windus} \emph
  {et~al.}}]{nwchem}%
  \BibitemOpen
  \bibfield  {author} {\bibinfo {author} {\bibfnamefont {M.}~\bibnamefont
  {Valiev}}, \bibinfo {author} {\bibfnamefont {E.~J.}\ \bibnamefont {Bylaska}},
  \bibinfo {author} {\bibfnamefont {N.}~\bibnamefont {Govind}}, \bibinfo
  {author} {\bibfnamefont {K.}~\bibnamefont {Kowalski}}, \bibinfo {author}
  {\bibfnamefont {T.~P.}\ \bibnamefont {Straatsma}}, \bibinfo {author}
  {\bibfnamefont {H.~J.}\ \bibnamefont {Van~Dam}}, \bibinfo {author}
  {\bibfnamefont {D.}~\bibnamefont {Wang}}, \bibinfo {author} {\bibfnamefont
  {J.}~\bibnamefont {Nieplocha}}, \bibinfo {author} {\bibfnamefont
  {E.}~\bibnamefont {Apra}}, \bibinfo {author} {\bibfnamefont {T.~L.}\
  \bibnamefont {Windus}},  \emph {et~al.},\ }\href@noop {} {\bibfield
  {journal} {\bibinfo  {journal} {Computer Physics Communications}\ }\textbf
  {\bibinfo {volume} {181}},\ \bibinfo {pages} {1477} (\bibinfo {year}
  {2010})}\BibitemShut {NoStop}%
\bibitem [{\citenamefont {Nafziger}\ and\ \citenamefont
  {Wasserman}(2014)}]{NW14}%
  \BibitemOpen
  \bibfield  {author} {\bibinfo {author} {\bibfnamefont {J.}~\bibnamefont
  {Nafziger}}\ and\ \bibinfo {author} {\bibfnamefont {A.}~\bibnamefont
  {Wasserman}},\ }\href {\doibase 10.1021/jp504058s} {\bibfield  {journal}
  {\bibinfo  {journal} {The Journal of Physical Chemistry A}\ }\textbf
  {\bibinfo {volume} {118}},\ \bibinfo {pages} {7623} (\bibinfo {year}
  {2014})}\BibitemShut {NoStop}%
\bibitem [{\citenamefont {Mosquera}\ and\ \citenamefont
  {Wasserman}(2013)}]{MW13}%
  \BibitemOpen
  \bibfield  {author} {\bibinfo {author} {\bibfnamefont {M.~A.}\ \bibnamefont
  {Mosquera}}\ and\ \bibinfo {author} {\bibfnamefont {A.}~\bibnamefont
  {Wasserman}},\ }\href@noop {} {\bibfield  {journal} {\bibinfo  {journal}
  {Mol. Phys.}\ }\textbf {\bibinfo {volume} {111}},\ \bibinfo {pages} {505}
  (\bibinfo {year} {2013})}\BibitemShut {NoStop}%
\bibitem [{\citenamefont {Wesolowski}, \citenamefont {Ellinger},\ and\
  \citenamefont {Weber}(1998)}]{WEW98}%
  \BibitemOpen
  \bibfield  {author} {\bibinfo {author} {\bibfnamefont {T.~A.}\ \bibnamefont
  {Wesolowski}}, \bibinfo {author} {\bibfnamefont {Y.}~\bibnamefont
  {Ellinger}}, \ and\ \bibinfo {author} {\bibfnamefont {J.}~\bibnamefont
  {Weber}},\ }\href@noop {} {\bibfield  {journal} {\bibinfo  {journal} {J.
  Chem. Phys.}\ }\textbf {\bibinfo {volume} {108}},\ \bibinfo {pages} {6078}
  (\bibinfo {year} {1998})}\BibitemShut {NoStop}%
\bibitem [{\citenamefont {Goodpaster}\ \emph {et~al.}(2010)\citenamefont
  {Goodpaster}, \citenamefont {Ananth}, \citenamefont {Manby},\ and\
  \citenamefont {Miller~{III}}}]{GAMM10}%
  \BibitemOpen
  \bibfield  {author} {\bibinfo {author} {\bibfnamefont {J.~D.}\ \bibnamefont
  {Goodpaster}}, \bibinfo {author} {\bibfnamefont {N.}~\bibnamefont {Ananth}},
  \bibinfo {author} {\bibfnamefont {F.~R.}\ \bibnamefont {Manby}}, \ and\
  \bibinfo {author} {\bibfnamefont {T.~F.}\ \bibnamefont {Miller~{III}}},\
  }\href@noop {} {\bibfield  {journal} {\bibinfo  {journal} {J. Chem. Phys.}\
  }\textbf {\bibinfo {volume} {133}},\ \bibinfo {pages} {084103} (\bibinfo
  {year} {2010})}\BibitemShut {NoStop}%
\bibitem [{\citenamefont {Nafziger}, \citenamefont {Wu},\ and\ \citenamefont
  {Wasserman}(2011)}]{NWW11}%
  \BibitemOpen
  \bibfield  {author} {\bibinfo {author} {\bibfnamefont {J.}~\bibnamefont
  {Nafziger}}, \bibinfo {author} {\bibfnamefont {Q.}~\bibnamefont {Wu}}, \ and\
  \bibinfo {author} {\bibfnamefont {A.}~\bibnamefont {Wasserman}},\ }\href
  {\doibase doi:10.1063/1.3667198} {\bibfield  {journal} {\bibinfo  {journal}
  {J. Chem. Phys.}\ }\textbf {\bibinfo {volume} {135}},\ \bibinfo {pages}
  {234101} (\bibinfo {year} {2011})}\BibitemShut {NoStop}%
\bibitem [{\citenamefont {Morales}\ and\ \citenamefont
  {Mart\'{i}nez}(2004)}]{MM04}%
  \BibitemOpen
  \bibfield  {author} {\bibinfo {author} {\bibfnamefont {J.}~\bibnamefont
  {Morales}}\ and\ \bibinfo {author} {\bibfnamefont {T.~J.}\ \bibnamefont
  {Mart\'{i}nez}},\ }\href {\doibase 10.1021/jp0369342} {\bibfield  {journal}
  {\bibinfo  {journal} {J. Phys. Chem. A}\ }\textbf {\bibinfo {volume} {108}},\
  \bibinfo {pages} {3076} (\bibinfo {year} {2004})}\BibitemShut {NoStop}%
\bibitem [{\citenamefont {Becke}(1982)}]{B82}%
  \BibitemOpen
  \bibfield  {author} {\bibinfo {author} {\bibfnamefont {A.~D.}\ \bibnamefont
  {Becke}},\ }\href@noop {} {\bibfield  {journal} {\bibinfo  {journal} {J.
  Chem. Phys.}\ }\textbf {\bibinfo {volume} {76}},\ \bibinfo {pages} {6037}
  (\bibinfo {year} {1982})}\BibitemShut {NoStop}%
\bibitem [{\citenamefont {Makmal}, \citenamefont {K\"{u}mmel},\ and\
  \citenamefont {Kronik}(2009)}]{MKK09}%
  \BibitemOpen
  \bibfield  {author} {\bibinfo {author} {\bibfnamefont {A.}~\bibnamefont
  {Makmal}}, \bibinfo {author} {\bibfnamefont {S.}~\bibnamefont {K\"{u}mmel}},
  \ and\ \bibinfo {author} {\bibfnamefont {L.}~\bibnamefont {Kronik}},\
  }\href@noop {} {\bibfield  {journal} {\bibinfo  {journal} {J. Chem. Theory
  Comput.}\ }\textbf {\bibinfo {volume} {5}},\ \bibinfo {pages} {1731}
  (\bibinfo {year} {2009})}\BibitemShut {NoStop}%
\bibitem [{\citenamefont {Kobus}, \citenamefont {Laaksonen},\ and\
  \citenamefont {Sundholm}(1996)}]{KLS96}%
  \BibitemOpen
  \bibfield  {author} {\bibinfo {author} {\bibfnamefont {J.}~\bibnamefont
  {Kobus}}, \bibinfo {author} {\bibfnamefont {L.}~\bibnamefont {Laaksonen}}, \
  and\ \bibinfo {author} {\bibfnamefont {D.}~\bibnamefont {Sundholm}},\
  }\href@noop {} {\bibfield  {journal} {\bibinfo  {journal} {Comput. Phys.
  Commun.}\ }\textbf {\bibinfo {volume} {98}},\ \bibinfo {pages} {346}
  (\bibinfo {year} {1996})}\BibitemShut {NoStop}%
\bibitem [{\citenamefont {Laaksonen}, \citenamefont {Pyykk\"{o}},\ and\
  \citenamefont {Sundholm}(1983)}]{LPS83}%
  \BibitemOpen
  \bibfield  {author} {\bibinfo {author} {\bibfnamefont {L.}~\bibnamefont
  {Laaksonen}}, \bibinfo {author} {\bibfnamefont {P.}~\bibnamefont
  {Pyykk\"{o}}}, \ and\ \bibinfo {author} {\bibfnamefont {D.}~\bibnamefont
  {Sundholm}},\ }\href@noop {} {\bibfield  {journal} {\bibinfo  {journal} {Int.
  J. Quant. Chem.}\ }\textbf {\bibinfo {volume} {23}},\ \bibinfo {pages} {309}
  (\bibinfo {year} {1983})}\BibitemShut {NoStop}%
\bibitem [{\citenamefont {Grabo}, \citenamefont {Kreibich},\ and\ \citenamefont
  {Gross}(1997)}]{GKG97}%
  \BibitemOpen
  \bibfield  {author} {\bibinfo {author} {\bibfnamefont {T.}~\bibnamefont
  {Grabo}}, \bibinfo {author} {\bibfnamefont {T.}~\bibnamefont {Kreibich}}, \
  and\ \bibinfo {author} {\bibfnamefont {E.~K.~U.}\ \bibnamefont {Gross}},\
  }\href@noop {} {\bibfield  {journal} {\bibinfo  {journal} {Mol. Eng}\
  }\textbf {\bibinfo {volume} {7}},\ \bibinfo {pages} {27} (\bibinfo {year}
  {1997})}\BibitemShut {NoStop}%
\bibitem [{\citenamefont {Marques}, \citenamefont {Oliveira},\ and\
  \citenamefont {Burnus}(2012)}]{libxc}%
  \BibitemOpen
  \bibfield  {author} {\bibinfo {author} {\bibfnamefont {M.~A.}\ \bibnamefont
  {Marques}}, \bibinfo {author} {\bibfnamefont {M.~J.}\ \bibnamefont
  {Oliveira}}, \ and\ \bibinfo {author} {\bibfnamefont {T.}~\bibnamefont
  {Burnus}},\ }\href@noop {} {\bibfield  {journal} {\bibinfo  {journal}
  {Comput. Phys. Commun.}\ }\textbf {\bibinfo {volume} {183}},\ \bibinfo
  {pages} {2272} (\bibinfo {year} {2012})}\BibitemShut {NoStop}%
\bibitem [{\citenamefont {Yang}, \citenamefont {Zhang},\ and\ \citenamefont
  {Ayers}(2000)}]{YZA00}%
  \BibitemOpen
  \bibfield  {author} {\bibinfo {author} {\bibfnamefont {W.}~\bibnamefont
  {Yang}}, \bibinfo {author} {\bibfnamefont {Y.}~\bibnamefont {Zhang}}, \ and\
  \bibinfo {author} {\bibfnamefont {P.~W.}\ \bibnamefont {Ayers}},\ }\href
  {\doibase 10.1103/PhysRevLett.84.5172} {\bibfield  {journal} {\bibinfo
  {journal} {Phys. Rev. Lett.}\ }\textbf {\bibinfo {volume} {84}},\ \bibinfo
  {pages} {5172} (\bibinfo {year} {2000})}\BibitemShut {NoStop}%
\bibitem [{\citenamefont {Perdew}\ \emph {et~al.}(1982)\citenamefont {Perdew},
  \citenamefont {Parr}, \citenamefont {Levy},\ and\ \citenamefont
  {Balduz~Jr}}]{PPLB82}%
  \BibitemOpen
  \bibfield  {author} {\bibinfo {author} {\bibfnamefont {J.~P.}\ \bibnamefont
  {Perdew}}, \bibinfo {author} {\bibfnamefont {R.~G.}\ \bibnamefont {Parr}},
  \bibinfo {author} {\bibfnamefont {M.}~\bibnamefont {Levy}}, \ and\ \bibinfo
  {author} {\bibfnamefont {J.~L.}\ \bibnamefont {Balduz~Jr}},\ }\href@noop {}
  {\bibfield  {journal} {\bibinfo  {journal} {Phys. Rev. Lett.}\ }\textbf
  {\bibinfo {volume} {49}},\ \bibinfo {pages} {1691} (\bibinfo {year}
  {1982})}\BibitemShut {NoStop}%
\bibitem [{\citenamefont {Dirac}(1930)}]{D30}%
  \BibitemOpen
  \bibfield  {author} {\bibinfo {author} {\bibfnamefont {P.~A.~M.}\
  \bibnamefont {Dirac}},\ }\href {\doibase 10.1017/S0305004100016108}
  {\bibfield  {journal} {\bibinfo  {journal} {Math. Proc. Cambridge}\ }\textbf
  {\bibinfo {volume} {26}},\ \bibinfo {pages} {376} (\bibinfo {year}
  {1930})}\BibitemShut {NoStop}%
\bibitem [{\citenamefont {Perdew}\ and\ \citenamefont {Wang}(1992)}]{PW92}%
  \BibitemOpen
  \bibfield  {author} {\bibinfo {author} {\bibfnamefont {J.~P.}\ \bibnamefont
  {Perdew}}\ and\ \bibinfo {author} {\bibfnamefont {Y.}~\bibnamefont {Wang}},\
  }\href@noop {} {\bibfield  {journal} {\bibinfo  {journal} {Phys. Rev. B}\
  }\textbf {\bibinfo {volume} {45}},\ \bibinfo {pages} {13244} (\bibinfo {year}
  {1992})}\BibitemShut {NoStop}%
\bibitem [{\citenamefont {Baer}, \citenamefont {Livshits},\ and\ \citenamefont
  {Salzner}(2010)}]{BLS10}%
  \BibitemOpen
  \bibfield  {author} {\bibinfo {author} {\bibfnamefont {R.}~\bibnamefont
  {Baer}}, \bibinfo {author} {\bibfnamefont {E.}~\bibnamefont {Livshits}}, \
  and\ \bibinfo {author} {\bibfnamefont {U.}~\bibnamefont {Salzner}},\
  }\href@noop {} {\bibfield  {journal} {\bibinfo  {journal} {Annu. Rev. Phys.
  Chem.}\ }\textbf {\bibinfo {volume} {61}},\ \bibinfo {pages} {85} (\bibinfo
  {year} {2010})}\BibitemShut {NoStop}%
\bibitem [{\citenamefont {Buijse}, \citenamefont {Baerends},\ and\
  \citenamefont {Snijders}(1989)}]{BBS89}%
  \BibitemOpen
  \bibfield  {author} {\bibinfo {author} {\bibfnamefont {M.~A.}\ \bibnamefont
  {Buijse}}, \bibinfo {author} {\bibfnamefont {E.~J.}\ \bibnamefont
  {Baerends}}, \ and\ \bibinfo {author} {\bibfnamefont {J.~G.}\ \bibnamefont
  {Snijders}},\ }\href@noop {} {\bibfield  {journal} {\bibinfo  {journal}
  {Phys. Rev. A}\ }\textbf {\bibinfo {volume} {40}},\ \bibinfo {pages} {4190}
  (\bibinfo {year} {1989})}\BibitemShut {NoStop}%
\bibitem [{\citenamefont {van Leeuwen}\ and\ \citenamefont
  {Baerends}(1994)}]{LB94a}%
  \BibitemOpen
  \bibfield  {author} {\bibinfo {author} {\bibfnamefont {R.}~\bibnamefont {van
  Leeuwen}}\ and\ \bibinfo {author} {\bibfnamefont {E.~J.}\ \bibnamefont
  {Baerends}},\ }\href@noop {} {\bibfield  {journal} {\bibinfo  {journal}
  {Phys. Rev. A}\ }\textbf {\bibinfo {volume} {49}},\ \bibinfo {pages} {2421}
  (\bibinfo {year} {1994})}\BibitemShut {NoStop}%
\bibitem [{\citenamefont {Gritsenko}, \citenamefont {van Leeuwen},\ and\
  \citenamefont {Baerends}(1995)}]{GLB95}%
  \BibitemOpen
  \bibfield  {author} {\bibinfo {author} {\bibfnamefont {O.~V.}\ \bibnamefont
  {Gritsenko}}, \bibinfo {author} {\bibfnamefont {R.}~\bibnamefont {van
  Leeuwen}}, \ and\ \bibinfo {author} {\bibfnamefont {E.~J.}\ \bibnamefont
  {Baerends}},\ }\href@noop {} {\bibfield  {journal} {\bibinfo  {journal}
  {Phys. Rev. A}\ }\textbf {\bibinfo {volume} {52}},\ \bibinfo {pages} {1870}
  (\bibinfo {year} {1995})}\BibitemShut {NoStop}%
\bibitem [{\citenamefont {Gritsenko}\ and\ \citenamefont
  {Baerends}(1996)}]{GB96}%
  \BibitemOpen
  \bibfield  {author} {\bibinfo {author} {\bibfnamefont {O.~V.}\ \bibnamefont
  {Gritsenko}}\ and\ \bibinfo {author} {\bibfnamefont {E.~J.}\ \bibnamefont
  {Baerends}},\ }\href@noop {} {\bibfield  {journal} {\bibinfo  {journal}
  {Phys. Rev. A}\ }\textbf {\bibinfo {volume} {54}},\ \bibinfo {pages} {1957}
  (\bibinfo {year} {1996})}\BibitemShut {NoStop}%
\bibitem [{\citenamefont {Gritsenko}\ and\ \citenamefont
  {Baerends}(1997)}]{GB97}%
  \BibitemOpen
  \bibfield  {author} {\bibinfo {author} {\bibfnamefont {O.~V.}\ \bibnamefont
  {Gritsenko}}\ and\ \bibinfo {author} {\bibfnamefont {E.~J.}\ \bibnamefont
  {Baerends}},\ }\href@noop {} {\bibfield  {journal} {\bibinfo  {journal}
  {Theor. Chem. Acc.}\ }\textbf {\bibinfo {volume} {96}},\ \bibinfo {pages}
  {44} (\bibinfo {year} {1997})}\BibitemShut {NoStop}%
\bibitem [{\citenamefont {Helbig}, \citenamefont {Tokatly},\ and\ \citenamefont
  {Rubio}(2009)}]{HTR09}%
  \BibitemOpen
  \bibfield  {author} {\bibinfo {author} {\bibfnamefont {N.}~\bibnamefont
  {Helbig}}, \bibinfo {author} {\bibfnamefont {I.~V.}\ \bibnamefont {Tokatly}},
  \ and\ \bibinfo {author} {\bibfnamefont {A.}~\bibnamefont {Rubio}},\
  }\href@noop {} {\bibfield  {journal} {\bibinfo  {journal} {J. Chem. Phys.}\
  }\textbf {\bibinfo {volume} {131}},\ \bibinfo {pages} {224105} (\bibinfo
  {year} {2009})}\BibitemShut {NoStop}%
\bibitem [{\citenamefont {Tempel}, \citenamefont {Mart\'{i}nez},\ and\
  \citenamefont {Maitra}(2009)}]{TMM09}%
  \BibitemOpen
  \bibfield  {author} {\bibinfo {author} {\bibfnamefont {D.~G.}\ \bibnamefont
  {Tempel}}, \bibinfo {author} {\bibfnamefont {T.~J.}\ \bibnamefont
  {Mart\'{i}nez}}, \ and\ \bibinfo {author} {\bibfnamefont {N.~T.}\
  \bibnamefont {Maitra}},\ }\href@noop {} {\bibfield  {journal} {\bibinfo
  {journal} {J. Chem. Theory Comput.}\ }\textbf {\bibinfo {volume} {5}},\
  \bibinfo {pages} {770} (\bibinfo {year} {2009})}\BibitemShut {NoStop}%
\bibitem [{\citenamefont {Elliott}\ \emph {et~al.}(2012)\citenamefont
  {Elliott}, \citenamefont {Fuks}, \citenamefont {Rubio},\ and\ \citenamefont
  {Maitra}}]{EFRM12}%
  \BibitemOpen
  \bibfield  {author} {\bibinfo {author} {\bibfnamefont {P.}~\bibnamefont
  {Elliott}}, \bibinfo {author} {\bibfnamefont {J.~I.}\ \bibnamefont {Fuks}},
  \bibinfo {author} {\bibfnamefont {A.}~\bibnamefont {Rubio}}, \ and\ \bibinfo
  {author} {\bibfnamefont {N.~T.}\ \bibnamefont {Maitra}},\ }\href {\doibase
  10.1103/PhysRevLett.109.266404} {\bibfield  {journal} {\bibinfo  {journal}
  {Phys. Rev. Lett.}\ }\textbf {\bibinfo {volume} {109}},\ \bibinfo {pages}
  {266404} (\bibinfo {year} {2012})}\BibitemShut {NoStop}%
\bibitem [{\citenamefont {Fuks}\ \emph {et~al.}(2013)\citenamefont {Fuks},
  \citenamefont {Elliott}, \citenamefont {Rubio},\ and\ \citenamefont
  {Maitra}}]{FERM13}%
  \BibitemOpen
  \bibfield  {author} {\bibinfo {author} {\bibfnamefont {J.~I.}\ \bibnamefont
  {Fuks}}, \bibinfo {author} {\bibfnamefont {P.}~\bibnamefont {Elliott}},
  \bibinfo {author} {\bibfnamefont {A.}~\bibnamefont {Rubio}}, \ and\ \bibinfo
  {author} {\bibfnamefont {N.~T.}\ \bibnamefont {Maitra}},\ }\href {\doibase
  10.1021/jz302099f} {\bibfield  {journal} {\bibinfo  {journal} {J. Phys. Chem.
  Lett.}\ }\textbf {\bibinfo {volume} {4}},\ \bibinfo {pages} {735} (\bibinfo
  {year} {2013})}\BibitemShut {NoStop}%
\bibitem [{\citenamefont {Hellgren}, \citenamefont {Rohr},\ and\ \citenamefont
  {Gross}(2012)}]{HRG2012}%
  \BibitemOpen
  \bibfield  {author} {\bibinfo {author} {\bibfnamefont {M.}~\bibnamefont
  {Hellgren}}, \bibinfo {author} {\bibfnamefont {D.~R.}\ \bibnamefont {Rohr}},
  \ and\ \bibinfo {author} {\bibfnamefont {E.}~\bibnamefont {Gross}},\
  }\href@noop {} {\bibfield  {journal} {\bibinfo  {journal} {The Journal of
  chemical physics}\ }\textbf {\bibinfo {volume} {136}},\ \bibinfo {pages}
  {034106} (\bibinfo {year} {2012})}\BibitemShut {NoStop}%
\end{thebibliography}%

\end{document}